\documentclass[reprint,superscriptaddress,amsmath,longbibliography,amssymb,prb]{revtex4-2}

\usepackage{graphicx}
\usepackage{dcolumn}
\usepackage{bm}
\usepackage{multirow}
\usepackage{array}
\usepackage{makecell}
\usepackage[usenames,dvipsnames]{xcolor}
\definecolor{nblue}{rgb}{0.0, 0.0, 1.0}
\definecolor{magenta}{rgb}{0.79, 0.08, 0.48}
\usepackage[colorlinks,linkcolor=nblue,urlcolor=blue,citecolor=blue,plainpages=false,pdfpagelabels,breaklinks]{hyperref}

\newcommand{\beq}{\begin{equation}}
\newcommand{\eeq}{\end{equation}}
\newcommand{\bea}{\begin{eqnarray}}
\newcommand{\eea}{\end{eqnarray}}

\begin{document}

\title{{Theoretical Prediction of High-Temperature Superconductivity in SrAuH$_3$ at Ambient Pressure}}
\author{Bin Li}\email[Electronic addresses: ]{libin@njupt.edu.cn}
\affiliation{School of Science, Nanjing University of Posts and Telecommunications, Nanjing 210023, China}
\affiliation{Jiangsu Provincial Engineering Research Center of Low Dimensional Physics and New Energy, Nanjing University of Posts and Telecommunications, Nanjing 210023, China}

\author{Cong Zhu}
\affiliation{College of Electronic and Optical Engineering, Nanjing University of Posts and Telecommunications, Nanjing 210023, China}

\author{Junjie Zhai}
\affiliation{School of Science, Nanjing University of Posts and Telecommunications, Nanjing 210023, China}
\affiliation{Key Laboratory of Dark Matter and Space Astronomy, Purple Mountain Observatory, Chinese Academy of Sciences, Nanjing 210023, China}

\author{Chuanhui Yin}
\affiliation{College of Electronic and Optical Engineering, Nanjing University of Posts and Telecommunications, Nanjing 210023, China}

\author{Yuxiang Fan}
\affiliation{School of Science, Nanjing University of Posts and Telecommunications, Nanjing 210023, China}

\author{Jie Cheng}
\affiliation{School of Science, Nanjing University of Posts and Telecommunications, Nanjing 210023, China}
\affiliation{Jiangsu Provincial Engineering Research Center of Low Dimensional Physics and New Energy, Nanjing University of Posts and Telecommunications, Nanjing 210023, China}

\author{Shengli Liu}\email[Electronic addresses: ]{liusl@njupt.edu.cn}
\affiliation{School of Science, Nanjing University of Posts and Telecommunications, Nanjing 210023, China}

\author{Zhixiang Shi}\email[Electronic addresses: ]{101200002@seu.edu.cn}
\affiliation{School of Physics, Southeast University, Nanjing 211189, China}

\date{\today}
\begin{abstract}
We present a comprehensive computational investigation of electron-phonon interactions in MXH$_3$ hydride compounds, where $M$ represents alkali and post-transition metals, and $X$ denotes 3$d$, 4$d$, and 5$d$ transition metals. Our density functional theory calculations identify 17 dynamically stable compounds. Notably, SrAuH$_3$ and SrZnH$_3$ emerge as theoretical ambient-pressure superconductors with predicted critical temperatures ($T_c$) exceeding 100 K. Analysis of the electronic structure reveals that the $X$ component dominates the density of states at the Fermi level, playing a crucial role in determining electron-phonon coupling strength and superconducting properties. We elucidate the underlying mechanisms governing these properties through detailed examination of the electronic and vibrational spectra. Our findings may challenge the prevailing notion that high-$T_c$ superconductivity in hydrides requires extreme pressures, potentially paving the way for practical applications. This study also provides valuable insights to guide future experimental efforts in the synthesis of ambient-pressure hydride superconductors.
\end{abstract}
\maketitle

\section{Introduction}

Hydrogen-based superconductors have emerged as a major focus in condensed matter physics, distinct from iron-based and nickel-based superconductors \cite{tokura1989superconducting,li2019superconductivity}. These materials, notable for their diverse structures, rich symmetry, and potential for room-temperature superconductivity, have sparked intense research aimed at reducing their stabilization pressure while maintaining high transition temperatures. This pursuit of stable high-temperature hydrogen-based superconductors under ambient or mild pressure conditions has become a cutting-edge topic in the field. The theoretical foundation for this research lies in Bardeen-Cooper-Schrieffer (BCS) theory \cite{PhysRev.108.1175}, which relates the superconducting critical temperature ($T_c$) to the Debye temperature. Metallic hydrogen, with its exceptionally high Debye temperature and strong electron-phonon coupling, has long been predicted to be a high $T_c$ superconductor \cite{wigner1935possibility,ashcroft1968metallic}, driving the exploration of hydrogen-rich compounds as potential high-temperature superconductors.

Recent decades have seen significant breakthroughs in hydride superconductors, driven by advancements in high-pressure experimental techniques \cite{RevModPhys.90.015007,buzea2004assembling,rotundu2013high} and crystal structure prediction methods \cite{WANG20122063,jansen2015conceptual,liu2012hybrid}. Binary hydrides like H$_3$S \cite{drozdov2015conventional}, LaH$_{10}$ \cite{drozdov2019superconductivity,somayazulu2019evidence}, and YH$_9$ \cite{drozdov2019superconductivity,wang2022synthesis} have been experimentally confirmed to exhibit $T_c$ above 200 K under high pressures. Research has expanded to ternary hydrogen-rich compounds, with theoretical predictions suggesting even higher $T_c$ values, such as 473 K for Li$_2$MgH$_{16}$ at 250 GPa \cite{PhysRevLett.123.097001}. Experimentally, (La, Ce)H$_{9,10}$ has demonstrated a $T_c$ of 176 K at 100 GPa \cite{PhysRevLett.128.167001}. Incorporating anharmonic corrections often significantly influences the minimum pressures required for dynamical stability. For instance, in BaSiH$_8$ and SrSiH$_8$, the stable pressure ranges are strongly affected \cite{Lucrezi2022,Lucrezi2023}. Similarly, the stability threshold is reduced to 77~GPa for LaBH$_8$ \cite{PhysRevB.106.134509} and to 129~GPa for LaH$_{10}$ \cite{Errea2020}. Ternary hydrides following the ABH$_8$ template \cite{PhysRevB.104.L020511,PhysRevLett.128.047001,PhysRevB.104.134501,PhysRevB.106.134509} have been experimentally realized, with LaBeH$_{8}$ exhibiting a $T_c$ of 110~K at 80~GPa \cite{PhysRevLett.130.266001}. Despite these advances, room-temperature superconductivity remains elusive, and the focus is shifting towards discovering materials that can achieve superconductivity at ambient or low pressures for practical applications. This evolving landscape encompasses polyhydrides, emphasizing the need for new hydrogen-based materials that balance high $T_c$ with feasible external pressure conditions.

Doping and elemental substitution in hydride compounds offer possible avenues for modifying atomic interactions and potentially lowering the pressure required for superconductivity. While introducing magnesium into methane produces $P4/nmm$-MgCH$_4$ with a $T_c$ of 120 K under high pressure~\cite{Tian_2015}, incorporating iron into H$_3$S to form Fe$_2$SH$_3$ drastically reduces the superconducting transition temperature to 0.3 K~\cite{zhang2016structure}. Recently, high-throughput computational screening predicted a metastable ambient-pressure hydride superconductor, Mg$_2$IrH$_6$, with a critical temperature of 160 K~\cite{PhysRevLett.132.166001}, comparable to the highest-temperature superconducting cuprates. These contrasting results highlight the complex interplay of elements in ternary hydrides and underscore the need for systematic exploration of new compositional and structural combinations. Such investigations are crucial for unlocking the potential of ternary hydrides as high-temperature superconductors that can operate at ambient pressure, a key requirement for practical applications. Future studies should focus on exploring a wider range of elemental combinations, investigating the role of crystal structure and chemical bonding, developing computational methods to predict stability and synthesizability, and experimental verification of computational predictions, with emphasis on ambient-pressure synthesis techniques.

In this study, we employ high-throughput computational methods to investigate potential superconductivity in cubic MXH$_3$ compounds under ambient pressure. Here, $M$ represents alkali and post-transition metals (Li, Na, Mg, Al, K, Ca, Ga, Rb, Sr, and In), while $X$ encompasses 3$\emph{d}$, 4$\emph{d}$, and 5$\emph{d}$ transition metals. Using first-principles calculations, we systematically examine the structural stability, electronic structure, phonon spectra, and superconducting properties of these 290 compounds. In the manuscript, we focus on the strontium hydrides: SrXH$_3$ (X = Au, Tc, Zn), which exhibit relatively high $T_c$. Our screening process prioritizes stability, a crucial factor in determining the feasibility of these materials for further study and potential applications. By focusing on ambient pressure conditions, our research aims to bridge the gap between high-$T_c$ hydrides and practical applications, advancing the field toward viable room-temperature superconductors. This comprehensive approach allows us to explore a wide range of compositions, potentially uncovering candidates for ambient-pressure, high-temperature superconductivity in the MXH$_3$ family.

\section{Methods}

We performed structure optimizations using the $\emph{ab initio}$ approach implemented in the Quantum Espresso (QE) package \cite{giannozzi2009quantum}. Charge density and wave function cutoff values were set to 600 Ry and 60 Ry, respectively. Self-consistent electron density calculations employed a 24$\times$24$\times$24 $k$-point mesh with a Gaussian smearing of 0.02 Ry. Following convergence tests, we calculated dynamical matrices and vibration potentials on a 8$\times$8$\times$8 $q$-point mesh using density functional perturbation theory (DFPT) \cite{baroni2001phonons}. An optimized tetrahedron method has been employed for the integration \cite{PhysRevB.89.094515}. We utilized Standard solid-state pseudopotentials \cite{prandini2018precision} throughout our calculations. Superconducting critical temperatures were estimated using the Allen-Dynes modified McMillan equation and numerical solutions of the Eliashberg equations \cite{allen1975transition}. For stable structures, we conducted high-accuracy electronic structure calculations using the full-potential linearized augmented plane wave (FP-LAPW) method implemented in the WIEN2K code \cite{blaha2020wien2k}, applying the generalized gradient approximation (GGA) \cite{perdew1996generalized} for exchange-correlation potentials. We visualized crystal structures using VESTA \cite{momma2011vesta} and Fermi surfaces using Fermisurfer \cite{kawamura2019fermisurfer}.
To estimate the superconducting $T_c$, we employed linear response calculations for \emph{e-ph} properties and applied the Allen-Dynes modified McMillan formula\cite{PhysRev.167.331,allen1975transition}:

\begin{equation}
{T_c} = f_1f_2\frac{{{\omega _{ln}}}}{{1.2}}{\rm{exp}}\left[ { - \frac{{1.04(1 + \lambda )}}{{\lambda  - {\mu ^*}(1 + 0.62\lambda )}}} \right],\\
\end{equation}
\begin{equation}
f_{1}=\left[1+\left(\lambda / \Lambda_{1}\right)^{3 / 2}\right]^{1 / 3}, \\
f_{2}=1+\frac{\left(\bar{\omega}_{2} / \omega_{ln}-1\right) \lambda^{2}}{\lambda^{2}+\Lambda_{2}^{2}},\\
\end{equation}
\begin{equation}
\Lambda_{1}=2.46\left(1+3.8 \mu^{*}\right), \\
\Lambda_{2}=1.82\left(1+6.3 \mu^{*}\right)\left(\bar{\omega}_{2} / \omega_{ln}\right),\\
\end{equation}

where $\omega_{ln}$ the logarithmically averaged phonon frequency, $\lambda$ the \emph{e-ph} coupling constant, and $\mu^{*}$ the Coulomb pseudopotential (set to 0.13). Factors $f_1$ and $f_2$ represent strong coupling and spectral function corrections, respectively, and depend on $\lambda$, $\mu^{*}$, $\omega_{ln}$, and the mean square frequency $\overline{\omega^2}$.

\section{Results and discussion }
\begin{figure*}
\begin{center}{\includegraphics[width=1\textwidth]{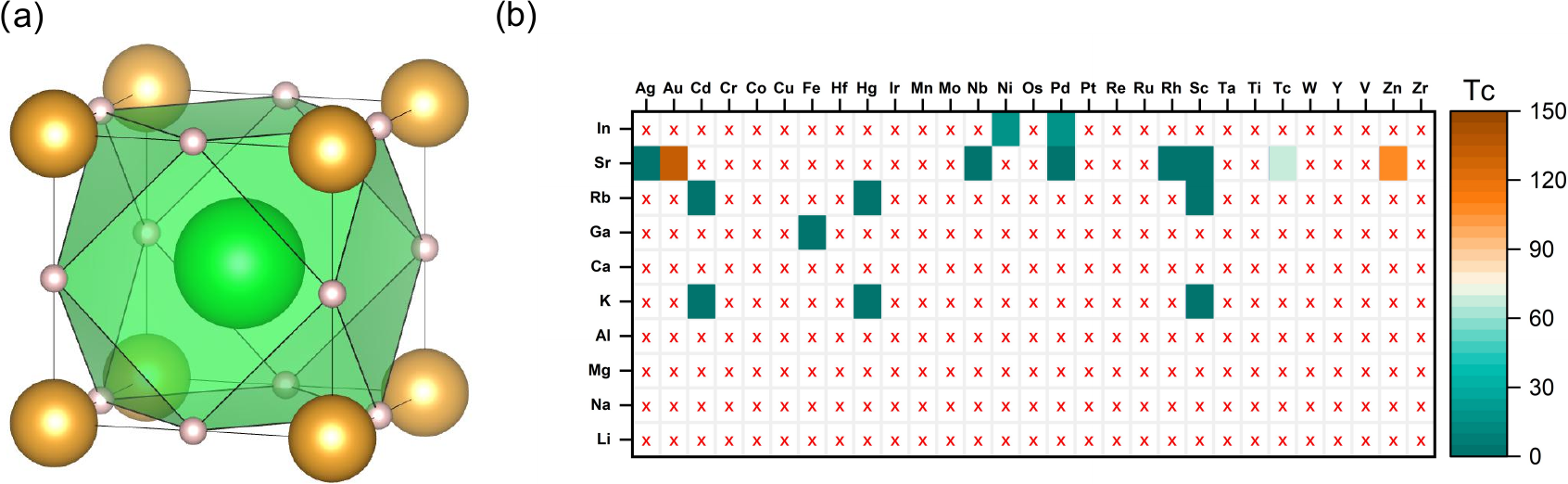}}
\caption{\label{fig1} (a) Crystal structure of SrAuH$_3$, in which the green, brown, and pink balls denote Sr, Au, and H atoms, respectively. (b) Superconducting critical temperature landscape of MXH$_3$ compounds. Heatmap representation of $T_c$ values for MXH$_3$ systems as a function of constituent metal atoms $X$ ($x$-axis) and $M$ ($y$-axis). Crosshatched cells (×) denote dynamically unstable phases as determined by phonon dispersion calculations.}
\end{center}
\end{figure*}

The crystal structure of SrAuH$_3$ (Fig. \ref{fig1}(a))crystallizes in the cubic space group $Pm\bar{3}m$ (No. 221). Sr, Au, and H atoms occupy 1$b$ (0.5, 0.5, 0.5), 1$a$ (0, 0, 0), and 3$d$ (0, 0, 0.5) Wyckoff positions, respectively, with a calculated lattice constant of 3.828 Å at ambient pressure. The detailed structural parameters and calculated superconducting properties for all 17 dynamically stable MXH$_3$ compounds are presented in Table S1 in Supplemental Material \cite{SupplementalMaterial}.  Figure~\ref{fig1}(b) presents a comprehensive heatmap of the calculated superconducting critical temperatures for the MXH$_3$ compounds. Our high-throughput calculations revealed eight dynamically stable structures with $M$ = Sr. Among these, SrAuH$_3$, SrTcH$_3$, and SrZnH$_3$ emerge as particularly  candidates, with Allen-Dynes $T_c$ predictions of 132 K, 69 K, and 107 K, respectively. The electron-phonon (\emph{e-ph}) coupling constants for these compounds are 2.025, 1.303, and 1.94, respectively, indicating strong coupling regimes. Notably, SrAuH$_3$ and SrZnH$_3$ exhibit the highest $T_c$s, surpassing the boiling point of liquid nitrogen. The logarithmic average phonon frequencies for these compounds are 806 K, 655 K, and 681 K, respectively. These results underscore the potential of Sr-based ternary hydrides as ambient-pressure high-$T_c$ superconductors and provide valuable insights into the interplay between electronic structure and lattice dynamics in these systems.

\begin{figure}
\begin{center}{\includegraphics[width=0.45\textwidth]{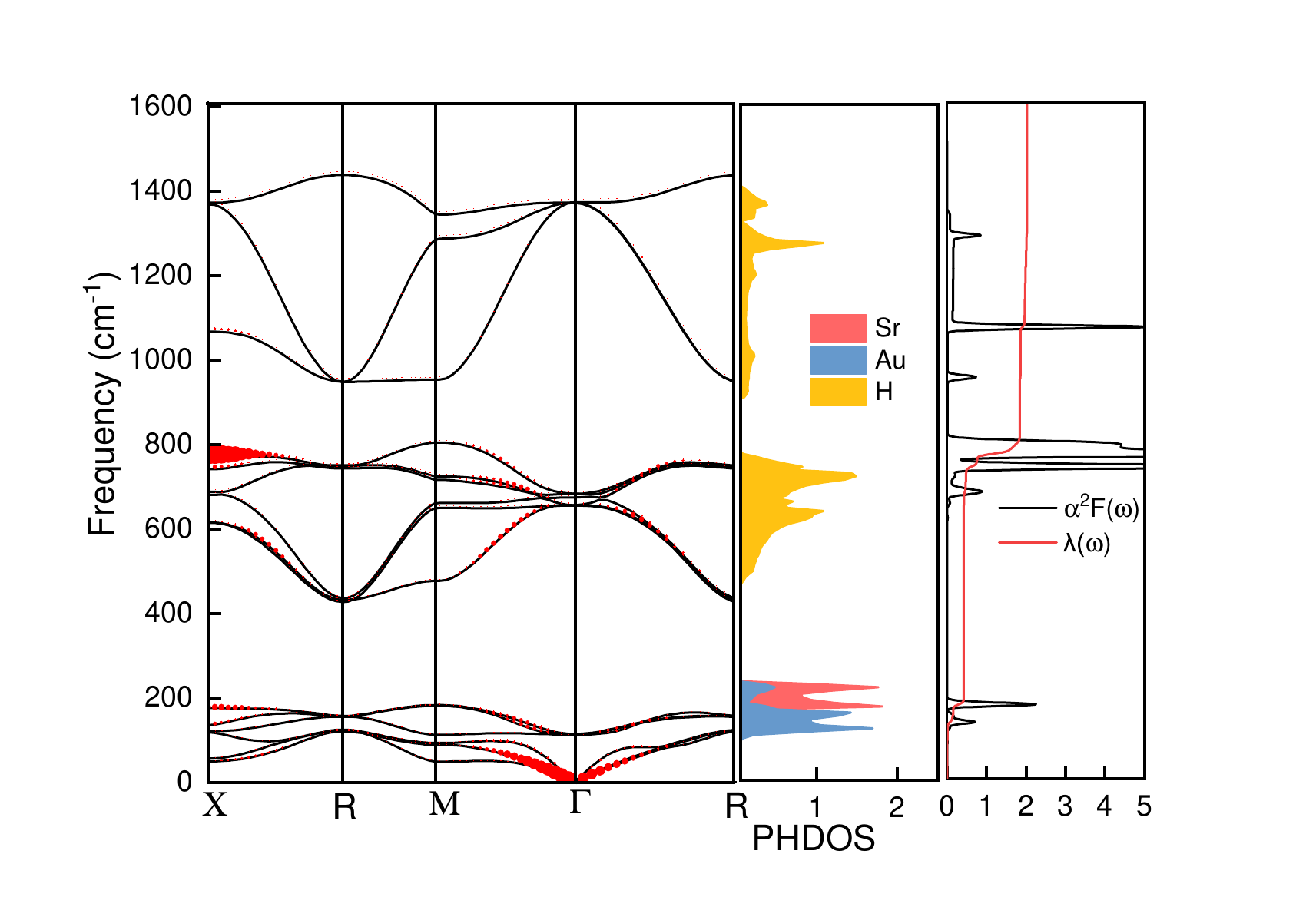}}
\caption{\label{fig2} 
Vibrational and electron-phonon coupling properties of SrAuH$_3$ at ambient pressure. Left panel: Phonon dispersion with electron-phonon coupling strength $\lambda_{\nu\mathbf{q}}$ (red circles, size proportional to coupling strength). Middle panel: Projected phonon density of states (PHDOS) for Sr (pink), Au (blue), and H (yellow). Right panel: Eliashberg spectral function $\alpha^2F(\omega)$ (black line, left axis) and frequency-dependent electron-phonon coupling $\lambda(\omega)$ 
}
\end{center}
\end{figure}

Figure~\ref{fig2} presents the phonon dispersions and projected phonon density of states (PHDOS) for SrAuH$_3$ at ambient pressure. The absence of imaginary frequencies along high-symmetry paths confirms dynamical stability. The highly dispersive top group of phonon spectra, peaking at $\sim$1420~cm$^{-1}$ at the $R$ point, is attributed to the hydrogen vibrations. Electron-phonon coupling coefficients $\lambda_{\nu\mathbf{q}}$ are overlaid on the dispersion curves, revealing strong couplings predominantly in acoustic and low-frequency optical modes near the $\Gamma$ point, possibly due to Au atom contributions. The Eliashberg function $\alpha^2F(\omega)$ and integrated $e$-ph coupling $\lambda(\omega)$ are shown in the right panels. The total $e$-ph coupling $\lambda = 2\int\alpha^2F(\omega)\omega^{-1}d\omega$ and logarithmically averaged phonon frequency $\omega_{ln} = \exp[2\lambda^{-1}\int d\omega\alpha^2F(\omega)\omega^{-1}\log\omega]$ are calculated. The coupling integral below 1200~cm$^{-1}$ accounts for the majority of the total coupling. Phonon spectra and densities of states for SrTcH$_3$ and SrZnH$_3$ are provided in the Supplemental Material \cite{SupplementalMaterial}. The primitive cell of SrAuH$_3$ contains 5 atoms, yielding 15 phonon bands: 3 acoustic and 12 optical branches, with irreducible representation expressed as 4$T_{1u}$ $\oplus$ 1$T_{2u}$. Low-frequency acoustic branches primarily arise from Sr and Au vibrations, while modes between 200 and 1400~cm$^{-1}$ are associated with H vibrations. The right panel of Figure ~\ref{fig2} shows the Eliashberg spectral function $\alpha^2F(\omega)$ and \emph{e-ph} coupling integration $\lambda(\omega)$. The total $\lambda$ is predominantly contributed by low- and medium-frequency vibrations.

\begin{figure}
\begin{center}{\includegraphics[width=0.45\textwidth]{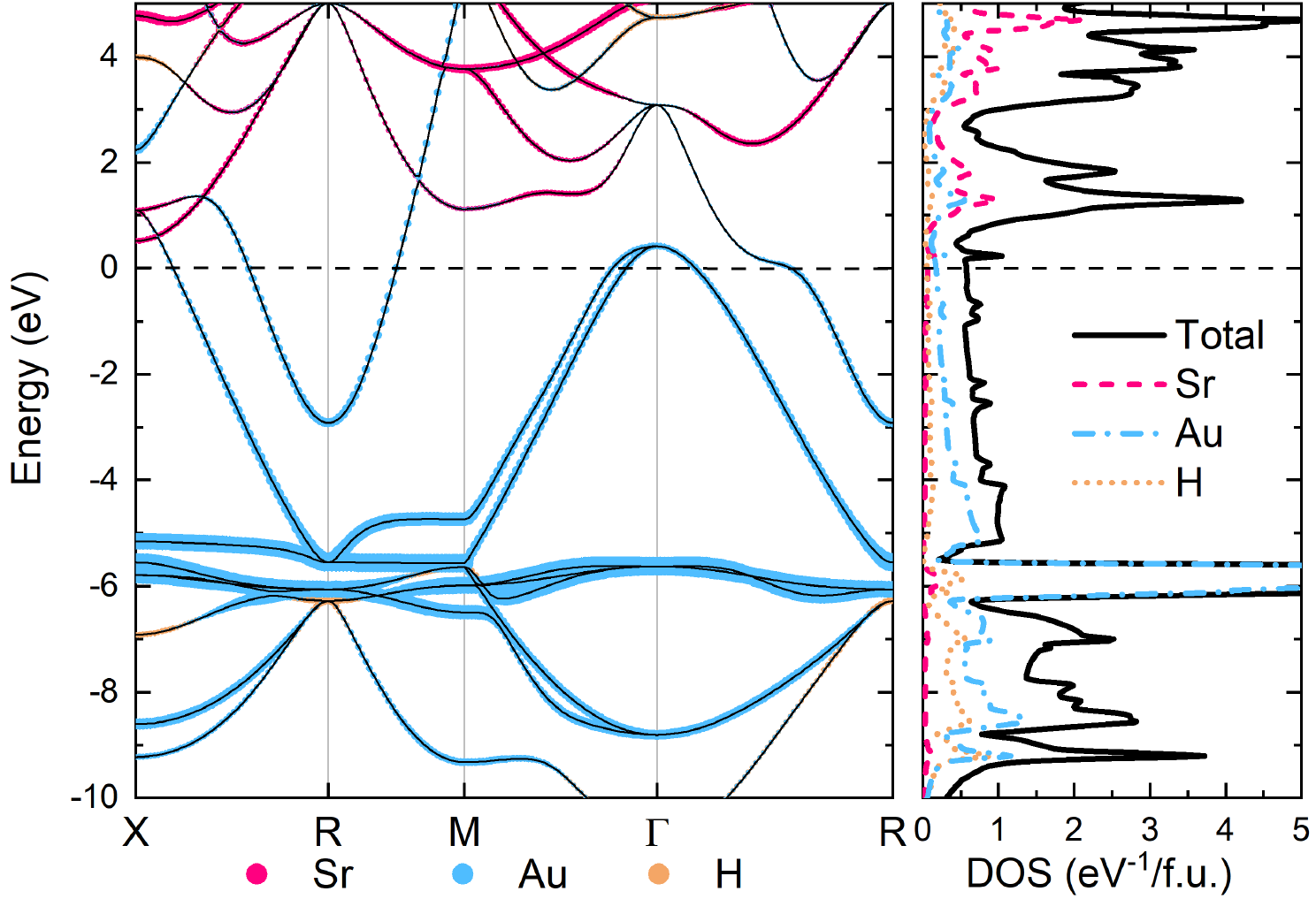}}
\caption{\label{fig3} 
Electronic structure of SrAuH$_3$ at ambient pressure. Left panel: Band structure along high-symmetry paths in the Brillouin zone. Right panel: Partial density of states (DOS) for Sr (pink), Au (blue), and H (yellow) atoms. The Fermi level is set to zero energy. DOS units are states/eV/formula unit.}
\end{center}
\end{figure}

The electronic band structure and atomic projected density of states (DOS) in eV$^{-1}$/formula unit (f.u.) for SrAuH$_3$ at ambient pressure are shown in Figure ~\ref{fig3}. The band structure exhibits metallic behavior, evidenced by multiple bands crossing the Fermi level ($E_F$). Near $E_F$, Au predominantly contributes to the DOS, with Au $5d$ orbitals dominating and forming a sharp van Hove singularity (vHs) approximately 0.2 eV above $E_F$. This vHs significantly enhances the DOS, potentially boosting superconductivity as per BCS theory. The high DOS at $E_F$, primarily from Au states, likely facilitates strong electron-phonon coupling, possibly explaining SrAuH$_3$'s high critical temperature. For comparison, electronic structures of SrTcH$_3$ and SrZnH$_3$ are provided in the Supplemental Material~\cite{SupplementalMaterial}. Analysis of the electronic band structures reveals that the $X$ atoms predominantly contribute to DOS near the Fermi level, underscoring the critical role of the $X$ elements in determining the electronic properties of these compounds.

\begin{figure}
\begin{center}{\includegraphics[width=0.45\textwidth]{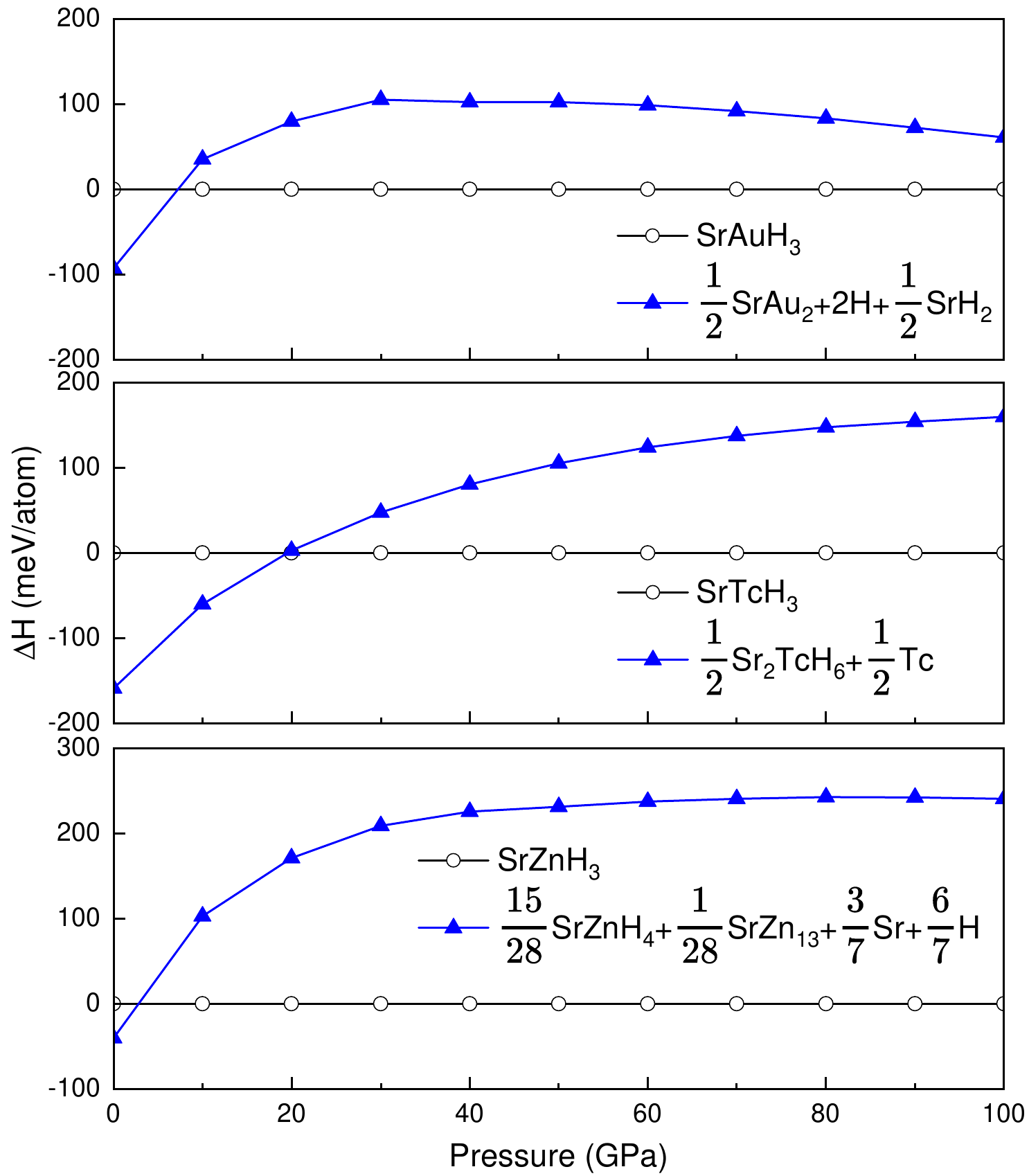}}
\caption{\label{fig4} Calculated enthalpy as a function of pressure for Sr--X--H structures relative to the $Pm\bar{3}m$ phase of SrXH$_3$ (X = Au, Tc, Zn).
}
\end{center}
\end{figure}
During experimental synthesis, materials can become trapped in local energy minima, known as metastable states, within the energy landscape. Figure~\ref{fig4} shows the calculated enthalpy as a function of pressure for Sr--X--H structures relative to the $Pm\bar{3}m$-SrXH$_3$. SrAuH$_3$ becomes thermodynamically stable above $\sim$7~GPa. The enthalpy difference $\Delta H$ = $\frac{1}{2}$ $H$({SrAu}$_2$+4{H} + {SrH}$_2$) -- $H$({SrAuH}$_3$) changes sign from negative to positive at this pressure. This crossover point delineates the stability regions of SrAuH$_3$ with respect to different decomposing phases, elucidating its pressure-driven formation mechanism. The low synthesis pressure of 7~GPa, achievable in multi-anvil apparatus, offers a possible route for experimental realization of this potential high-$T_c$ superconductor. 
Similarly, SrTcH$_3$ and SrZnH$_3$ undergo phase transitions at 20~GPa and 3~GPa, respectively, with SrTcH$_3$ possibly decomposed into Sr$_2$TcH$_6$ and Tc, and SrZnH$_3$ decomposed into SrZnH$_4$, SrZn$_{13}$, Sr and H.

\begin{figure}
\begin{center}{\includegraphics[width=0.5\textwidth]{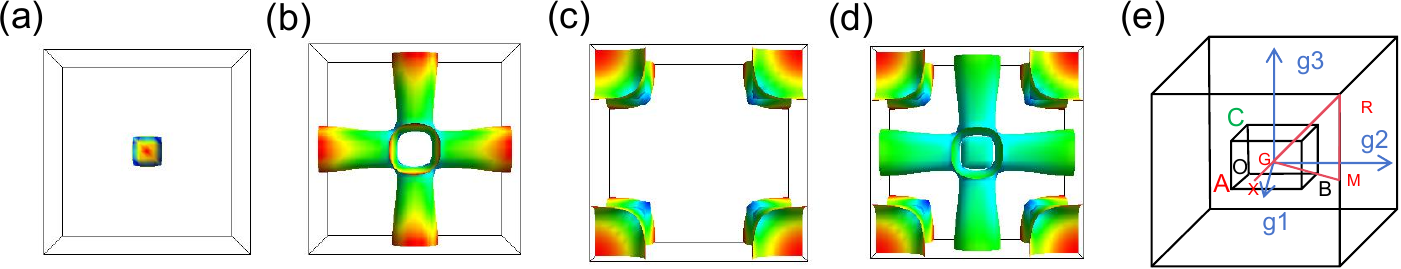}}
\caption{\label{fig5} 
Fermi surface of SrAuH$_3$ at ambient pressure. (a-c) Individual sheets of the Fermi surface. (d) Complete Fermi surface. Color scale indicates the Fermi velocity magnitude. (e) First Brillouin zone with high-symmetry points labeled.
}
\end{center}
\end{figure}

Detailed visualization of the Fermi surfaces for SrAuH$_3$ at ambient pressure are shown in Figure ~\ref{fig5}. The surfaces are color-coded to represent the Fermi velocity distribution, with the spectrum transitioning from blue to red as velocity increases. This complex Fermi surface topology comprises three distinct components: (1) a nearly spherical hole-like sheet centered at the $\Gamma$ point, dominating the Brillouin zone center; (2) cross-shaped sheets extending along the principal axes, indicative of significant band dispersion in these directions; and (3) eight electron-like pockets located at the Brillouin zone corners ($R$ points). The hole-like character of the central sphere, combined with the electron-like nature of the corner pockets, suggests a compensated metallic behavior. Analysis of the band structure near the Fermi level confirms the predominantly electronic nature of the Fermi surface, with the Au $5d$ contributing significantly to the density of states at $E_F$. The observed Fermi surface complexity, particularly the presence of multiple sheets with varying electron and hole characters, may enhance electron-phonon coupling through increased scattering channels. Furthermore, the high Fermi velocities, indicated by the red regions, imply large electronic bandwidths, which could contribute to the predicted high critical temperature  in this system. These features collectively suggest that SrAuH$_3$ possesses the electronic characteristics often associated with high-temperature superconductors, warranting further investigation into its superconducting properties and potential applications.

\section{Conclusion}

In conclusion, we present a comprehensive high-throughput investigation of superconductivity in cubic ternary hydrides MXH$_3$ at ambient pressure. Our analysis identifies 17 dynamically stable strontium hydrides, with SrAuH$_3$, SrTcH$_3$ and SrZnH$_3$ exhibiting notably high superconducting transition temperatures of 132~K, 69~K and 107~K, respectively. The exceptional $T_c$ of SrXH$_3$ suggests a significant influence of $X$ atoms on superconductivity in strontium hydrides. The predicted synthesis pressures for SrXH$_3$ compounds fall within experimentally accessible ranges ($\leq$ 20 GPa), providing possible avenues for future high-pressure synthesis experiments. These findings not only reveal a rich landscape of potential high-$T_c$ superconductors within the MXH$_3$ family at ambient pressure but also demonstrate an effective strategy for exploring conventional superconductivity in ternary hydrides through distinctive structural motifs. This study provides a foundation for future investigations into ambient-pressure, high-$T_c$ ternary hydride superconductors, potentially catalyzing new avenues in superconductor research and materials design.

\section*{Acknowledgements}
This work is supported by the National Natural Science Foundation of China (Grants No. 12374135, 12175107), and Nanjing University of Posts and Telecommunications Foundation (NUPTSF) (Grants No. NY219087, NY220038) and the Hua Li Talents Program of Nanjing University of Posts and Telecommunications. Some of the calculations were performed on the supercomputer in the Big Data Computing Center (BDCC) of Southeast University.\\
%

\end{document}